\documentclass[a4paper,12pt]{article}
\usepackage{amsmath}
\usepackage{amsfonts}
\usepackage{amssymb}
\usepackage{amsthm}
\usepackage{mathrsfs}
\usepackage{bm}
\usepackage{graphics,color}
\newtheorem{theorem}{Theorem}
\newtheorem{lemma}{Lemma}
\newtheorem{cor}{Corollary}

\theoremstyle{remark}
\newtheorem{remark}{Remark}
\theoremstyle{definition}
\newtheorem{defn}{Definition}

\newcommand{\cir}{|\mbox{Circ}\rangle_n}

\newcommand{\ellip}{|\mbox{Elliptic}(\theta)\rangle}

\newcommand{\ex}[1]{\mathbb{E}\left\{ #1 \right\}}
\newcommand{\pr}[1]{\mathbb{P}\left\{ #1 \right\}}
\newcommand{\hf}[1][1]{\frac{#1}{2}}

\newcommand{\di}{\,\mathrm{d}}
\newcommand{\ii}{\mathrm{i}}
\renewcommand{\vec}[1]{\mbox{\boldmath$#1$}}

\begin{document}
\title{On the divine clockwork: the spectral gap for the correspondence limit of the Nelson diffusion generator for the atomic elliptic state.}
\author{Richard Durran $\quad$  Andrew Neate \\ Aubrey Truman $\quad$  Feng-Yu Wang\\
\small Department of Mathematics, Swansea University, \\
\small Singleton Park, Swansea, SA2 8PP, Wales, UK.}
\maketitle
\begin{abstract}
The correspondence limit of the atomic elliptic state in three dimensions is discussed in terms of Nelson's stochastic mechanics. In previous work we have  shown that this approach leads to a limiting Nelson diffusion and here we discuss in detail the invariant measure for this  process and show that it is concentrated on the Kepler ellipse in the plane $z=0$. We then show that the limiting Nelson diffusion generator has a spectral gap; thereby proving  that in the infinite time limit the density for the limiting Nelson diffusion will converge to its invariant measure. 
We also include a  summary of the Cheeger and Poincar\'{e} inequalities both of which are used in our proof of the existence of the spectral gap.
\end{abstract}

\section{Introduction}
Nelson's stochastic mechanics \cite{Nelson3, Nelson1, Nelson2} encapsulates the dynamical content of the Schr\"{o}dinger equation. It describes the motion of a quantum particle of mass $m$ using a diffusion process driven by a Brownian motion with diffusion coefficient $\hbar/2m$. This diffusion process obeys a stochastic generalisation of Newton's second law of motion known as the Nelson-Newton law;  the mass times the stochastic acceleration equals the force. Considering the dynamics of such a Nelson diffusion leads to a set of equations which are equivalent to the Schr\"{o}dinger equation.
 Consequently, this approach allows the development of the theory of quantum mechanics without the need to entirely abandon the classical world. The fundamental aspects of quantum theory such as  the uncertainty principle are automatically built into this stochastic framework. Moreover, Nelson's theory gives a natural setting in which to consider the correspondence limit transition between the quantum and classical worlds.

In \cite{Divine} we investigated the Nelson diffusion for the atomic elliptic state in the Coulomb potential. We derived a limiting Nelson diffusion according to Bohr's correspondence principle and showed that in a two dimensional setting, the trajectories of this limiting diffusion converged to those of a particle describing Kepler's ellipse at a rate consistent with Kepler's laws of planetary motion. This solved a long standing problem in quantum mechanics in the two dimensional case. In this paper we will show how these results can be extended to the three dimensional case. Our main result is to show that there exists a spectral gap for the generator of the limiting Nelson diffusion for the atomic elliptic state. It follows that in the infinite time limit this process will converge to its invariant measure. Moreover, we will show that the invariant measure is, in the limit, concentrated in a neighbourhood of the Kepler ellipse in the plane $z=0$.

We begin by recapping the construction of the limiting Nelson diffusion from \cite{Divine}; in Section 2 we give an account of Nelson's stochastic mechanics before we introduce the atomic elliptic state in Section 3 and  discuss the correspondence limit  in Section 4.  In Section 5 we discuss the invariant measure for the limiting Nelson diffusion and show that it is the physically correct measure concentrated on the Kepler ellipse in the plane $z=0$. In Section 6 we discuss Cheeger's inequality and the Poincar\'{e} inequality which we use to prove the existence of the spectral gap in Section 7.

\section{Nelson's stochastic mechanics}
We begin with a brief look at the dynamics of diffusion processes \cite{Nelson1}. Consider a particle of mass $m$ diffusing through $\mathbb{R}^d$ according to the It\^{o} stochastic differential  equation,
\begin{equation}\label{diffusion1}
\di \bm{X}(t) = \bm{b}(\bm{X}(t),t)\di t+\epsilon\di\bm{B}(t),\quad t>0,
\end{equation}
where $\bm{B} = (B_1,B_2,\ldots,B_d)$ is a $d$-dimensional Brownian motion with normalisation $\ex{B_i(t)B_j(s)} = \delta_{ij}\min(t,s)$ for $ i,j = 1,2,\ldots,d.$
It follows from It\^{o}'s formula that the density  $\rho(\bm{x},t)$ defined for all Borel sets $A\subset \mathbb{R}^d$ by,
\[\pr{\bm{X}(t)\in A} = \int_{\bm{y}\in A} \rho (\bm{y},t)\di \bm{y},\]
 satisfies the forward Kolmogorov equation,
\begin{equation}\label{FK1}
\frac{\partial \rho}{\partial t} = \nabla\cdot\left(\frac{\epsilon^2}{2}\nabla \rho - \bm{b} \rho\right) = \mathcal{G}^*\rho,\qquad \rho(\bm{x},0) = \rho_0(\bm{x})
\end{equation}
where $\mathcal{G}^*$ is the $L^2$ adjoint of the generator $\mathcal{G}$ for the  diffusion $\bm{X}$:
\begin{equation}\label{G1}
\mathcal{G} = \frac{\epsilon^2}{2}\Delta + \bm{b}\cdot\nabla.
\end{equation}

The paths of  $\bm{X}$ are nowhere differentiable with probability 1, and so to discuss the kinematics of diffusions, Nelson introduced the mean forward and backward derivatives as conditional expectations,
\[D_{\pm} f(\bm{X}(t),t) := \lim\limits_{h\downarrow 0} \ex{\left.\frac{f(\bm{X}(t\pm h),t\pm h)-f(\bm{X}(t),t)}{\pm h}\right|\bm{X}(t)}.\]
The mean forward and backward drifts $\bm{b}_{\pm}$ are defined by,
\begin{equation}\label{FD1}
\bm{b}_+(\bm{X}(t),t):=D_+ \bm{X}(t)  = \bm{b}(\bm{X}(t),t),
\end{equation}
and
\begin{equation}\label{BD1}
 \bm{b}_-(\bm{X}(t),t):=D_- \bm{X}(t) = \bm{b}(\bm{X}(t),t) - \epsilon^2\nabla\ln\rho(\bm{X}(t),t).
 \end{equation}
The osmotic and current velocities $\bm{u}$ and $\bm{v}$ are defined by,
\begin{equation}\label{OV1}
\bm{u} := \frac{1}{2}(\bm{b}_+-\bm{b}_-),\quad\bm{v} := \frac{1}{2}(\bm{b}_++\bm{b}_-).
\end{equation}

The dynamics are introduced through the stochastic acceleration,
\[\bm{a}(\bm{X}(t),t) = \hf\left(D_+D_-+D_-D_+\right)\bm{X}(t),\]
and it follows from It\^{o}'s formula that,
\begin{equation}\label{SA2}
\bm{a}(\bm{X}(t),t) = \left(\frac{\partial \bm{v}}{\partial t} +(\bm{v}\cdot\nabla)\bm{v}-(\bm{u}\cdot\nabla)\bm{u} -\hf[\epsilon^2]\Delta \bm{u}\right)(\bm{X}(t),t).
\end{equation}

We will now discuss the connection between the dynamics of diffusion processes and quantum mechanics.
Consider a particle of mass $m$ in the quantum state $\psi = \psi(\bm{x},t)$ satisfying the Schr\"{o}dinger equation,
\begin{equation}\label{SE1}
i\hbar\frac{\partial \psi}{\partial t} =-\frac{\hbar^2}{2 m}\Delta  \psi +V\psi,\quad\bm{x}\in\mathbb{R}^d,\quad t>0.
\end{equation}
We can write this wave function in the form
$\psi = \exp(R+iS)$
where $R$ and $S$ are real valued. Nelson's theory states that such a particle can be considered to diffuse through $\mathbb{R}^d$ according to the It\^{o} stochastic differential equation (\ref{diffusion1}) where we choose $\bm{b} = \epsilon^2\nabla(R+S)$ and $\epsilon = \sqrt{\hbar/m}$. We refer to such a process as a Nelson diffusion.
This can be seen by writing the Schr\"{o}dinger equation (\ref{SE1}) in the form,
\begin{equation}\label{SE2}
i\hbar \psi^*\frac{\partial\psi}{\partial t} = -\frac{\hbar^2}{2m}\psi^*\Delta\psi + V|\psi|^2,
\end{equation}
where $\psi^*$ is the complex conjugate of $\psi$. Equating imaginary parts gives the continuity equation for the probability current $\bm{j}$,
\begin{equation}\label{CE1}
\frac{\partial\rho}{\partial t} + \nabla\cdot \bm{j} = 0,\qquad \bm{j}:= \frac{\epsilon^2}{2 i}(\psi^*\nabla\psi-\psi\nabla\psi^*) = \epsilon^2\rho\nabla S,
\end{equation}
where  $\rho$ is the particle density $\rho := |\psi|^2 = \exp(2R).$
The continuity equation (\ref{CE1}) can be written as,
\[
\frac{\partial \rho}{\partial t}  = \nabla\cdot\left(\hf[\epsilon^2]\nabla \exp({2R})-\epsilon^2 \exp(2R)\nabla(R+S)\right),\]
which is the forward Kolmogorov equation (\ref{FK1}) for the Nelson diffusion with drift $\bm{b} = \epsilon^2\nabla(R+S)$ and density $\rho = \exp(2R)$. Therefore, we see that the Nelson theory is consistent with Born's probabilistic interpretation for the quantum mechanical particle density $\rho$.

It now  follows from equations (\ref{FD1}), (\ref{BD1}) and (\ref{OV1}) that,
\[ \bm{b}_+=\epsilon^2\nabla(R+S),\quad \bm{b}_-=  \epsilon^2\nabla(S-R),\quad \bm{u}= \epsilon^2\nabla R,\quad\bm{v} =\epsilon^2\nabla S.\]
Moreover, from equation (\ref{SA2}) and the real part of equation (\ref{SE2}) we obtain the Nelson-Newton law,
\begin{equation}\label{NelsonNewton}
m \bm{a}(\bm{X}(t),t) = -\nabla V(\bm{X}(t),t).
\end{equation}
This is the dynamical content of the Schr\"{o}dinger equation expressed in terms of the Nelson diffusion.

Since $\psi$ satisfies a complex heat equation, it follows that
$\mathcal{S} = -i\epsilon^2\ln\psi$
must satisfy a complex Hamilton Jacobi equation with viscosity $i\epsilon^2$,
\begin{equation}\label{HJ1}\frac{\partial \mathcal{S}}{\partial t} +\hf |\nabla\mathcal{S}|^2 +\frac{V}{m} = \hf[i\epsilon^2]\Delta \mathcal{S}.
\end{equation}
Hence, defining $\bm{Z}$ by a complex Hopf-Cole transformation,
\begin{equation}\label{Z1}\bm{Z}= -i\epsilon^2 \nabla\ln\psi = \epsilon^2\nabla(S-iR),
\end{equation}
we obtain the complex Burgers equation,
\begin{equation}\label{BE1} \frac{\partial \bm{Z}}{\partial t}+(\bm{Z}\cdot \nabla)\bm{Z}+\frac{\nabla V}{m} =
\hf[i\epsilon^2]\Delta \bm{Z},\end{equation}
where the drift of the Nelson diffusion is given by
\[\bm{b} = \mathrm{Re}( \bm{Z}) - \mathrm{Im} (\bm{Z}).\]

As is well known, equating real and imaginary parts of the complex Burgers equation (\ref{BE1}) gives the Madelung fluid equation and its continuity equation for the current velocity $\bm{v} = \epsilon^2\nabla S$,
\begin{equation}\label{MF1}\frac{\partial \bm{v}}{\partial t} + (\bm{v}\cdot\nabla)\bm{v}+\nabla V = \hf[\epsilon^4]\nabla\left(\frac{\Delta \rho^{1/2}}{\rho^{1/2}}\right),\qquad
\frac{\partial\rho}{\partial t}+\nabla\cdot(\rho\bm{v})=0.\end{equation}

When  $\psi$ is a stationary state we can write,
\[\psi(\bm{x},t) =\psi_E(\bm{x})\exp(-iEt/\hbar),\quad t>0,\]
and then $\psi_E(\bm{x}) = \exp(R+iS)$ which satisfies the stationary Schr\"{o}dinger equation,
\begin{equation}\label{SSE1}-\frac{\hbar^2}{2m}\Delta\psi_E+V\psi_E = E\psi_E.\end{equation}
Then 
$\bm{Z} $ satisfies,
\begin{equation}\label{SBE1}\hf\bm{Z}^2+\frac{V}{m}-\hf[i\epsilon^2] \nabla\cdot \bm{Z} = E.\end{equation}
Moreover, for a stationary state, $\rho = \exp(2R)$ is the invariant density of the quantum system, and so is also  an invariant measure for the Nelson diffusion.

\section{The atomic elliptic state}
We begin  with the atomic circular state  $\cir$  for a particle of unit mass given by,
\[\langle \bm{x}\cir = \Psi_{n,n-1,n-1}(\bm{x}),\]
where $\vec{x} = (x,y,z)$ and  $\Psi_{n,l,m}$ (with  $l = 0,1,\ldots,n-1,$ $|m|\le l$ and $m,n \in\mathbb{N} $) is the nodal Schr\"{o}dinger wave function  for the Hamiltonian of the hydrogen atom,
\[
H(\vec{p},\vec{q}) = \hf[\vec{p}^2]-\frac{\mu}{|\vec{q}|},
\]
with
$[q_k,p_l] = \ii \hbar \delta_{kl}$ for $k,l = 1,2,3.$ We choose suitable units so that $\hbar = 1$ and $\mu=1$.
Then, for the orbital angular momentum $\vec{L}$, where
\[\bm{L} =(L_1,L_2,L_3)= \bm{q}\bm{\times} \bm{p},\qquad \bm{L}^2 = (L_1^2+L_2^2+L_3^2),\]
we have the eigenvalue relations,
\[\bm{L}^2 \Psi_{n,l,m} = l(l+1)\Psi_{n,l,m},\quad L_3 \Psi_{n,l,m} = m\Psi_{n,l,m},\quad H \Psi_{n,l,m} = -\frac{1}{2 n^2}\Psi_{n,l,m}.
 \]
 The atomic elliptic state is then given by,
\[\ellip_n = \exp(-\ii\theta A_2) \cir,
\]
where,
\[\vec{A} = (A_1,A_2,A_3) = \frac{1}{\sqrt{-2E}} \left(\hf[(\vec{p}\bm{\times} \vec{L}-\vec{L}\bm{\times} \vec{p})]-\frac{\vec{q}}{|\vec{q}|}\right),\]
is the Hamilton-Lenz-Runge vector.

Following \cite{Lena},
to within a multiplicative constant (after reinstating $\hbar$ and the force constant $\mu$),
\begin{equation}\label{AES1}
\psi_{\epsilon,n}(\vec{x}):=\langle x |\mbox{Elliptic}(\theta)\rangle_n = \exp\left(-\frac{n\mu }{\lambda^2}|\vec{x}|\right) \mathcal{L}_{n-1}(n\nu),-\end{equation}
where
\begin{equation}\label{NU}
\nu = \frac{\mu}{\lambda^2}\left(|\vec{x}| - \frac{x}{e} - \frac{\ii y\sqrt{1-e^2}}{e}\right),
\end{equation}
with $\vec{x} = (x,y,z)$, $E_n = -\mu^2/(2\lambda^2)$, $\epsilon^2=\hbar$,  $\lambda = n \epsilon^2$
and $\mathcal{L}_{n-1}$ a Laguerre polynomial.
This wave function is localised on the ellipse with eccentricity $e = |\sin\theta|$ given by,
\[\mathcal{E}_e = \left\{(x,y,z):\quad\frac{(x-a e)^2}{a^2}+\frac{y^2}{a^2(1-e^2)} =1,\quad z=0\right\},\]
where $a = \lambda^2/\mu$.  We call $\mathcal{E}_e$ the Kepler ellipse. 
For integer $n$ this wave function  satisfies the stationary Schr\"{o}dinger equation,
\begin{equation}\label{SSE2}
-\hf\epsilon^4\Delta\psi_{\epsilon,n}(\vec{x})-\frac{\mu}{|\vec{x}|}\psi_{\epsilon,n}(\vec{x}) = E_n\psi_{\epsilon,n}(\vec{x}).
\end{equation}
\begin{remark}
Throughout this paper $e$ will refer to the eccentricity of the Kepler ellipse. When we use the exponential function we shall always write $\exp(x)$ rather than $e^x$ to avoid confusion.
\end{remark}

\section{The correspondence limit}
\subsection{The limiting state and limiting Nelson diffusion}
Let the atomic elliptic state from equation (\ref{AES1}) be written in the form,
\[\psi_{\epsilon,n}(\bm{x}) = \exp(R_{\epsilon,n}+iS_{\epsilon,n}).\]
This stationary state has a Nelson diffusion  $\bm{X}_{\epsilon,n}(t)$ given by,
\[\di \bm{X}_{\epsilon,n}(t) = \bm{b}_{\epsilon,n}(\bm{X}_{\epsilon,n}(t))\di t+\epsilon\di\bm{B}(t),\]
where $\bm{b}_{\epsilon,n} = \epsilon^2\nabla(R_{\epsilon,n}+S_{\epsilon,n})$. For integer $n$, the diffusion $\bm{X}_{\epsilon,n}$ satisfies the Nelson-Newton law (\ref{NelsonNewton}).

We now take the correspondence limit of $\bm{X}_{\epsilon,n}$ by considering the drift $\bm{b}_{\epsilon,n}$ as $n\rightarrow\infty$ while $\epsilon\rightarrow 0$ with $\lambda=n\epsilon^2$ fixed. This gives a new diffusion $\bm{X}_{\epsilon}$ such that,
\[\di \bm{X}_{\epsilon}(t) = \bm{b}(\bm{X}_{\epsilon}(t))\di t+\epsilon\di\bm{B}(t),\]
where
\[\bm{b} = \lim\limits_{\genfrac{}{}{0pt}{}{n\uparrow\infty, \epsilon\downarrow 0}{\lambda = n\epsilon^2}}
\epsilon^2\nabla(R_{\epsilon,n}+S_{\epsilon,n}).\]
 We will refer to $\bm{X}_{\epsilon}$ as the limiting Nelson diffusion.

For this we use the vector $\bm{Z}_{\epsilon,n} = -i\epsilon^2\nabla\ln\psi_{\epsilon,n}$ introduced in equation (\ref{Z1}). In Cartesians,
\begin{equation}\label{Z2}
 \bm{Z}_{\epsilon,n}(\bm{x}) = \frac{\ii\mu}{\lambda }\left(1
  - \frac{\mathcal{L}'_{n-1}(n\nu)}{\mathcal{L}_{n-1}(n\nu)}\right)
   \frac{ \bm{x}}{|\bm{x}|}+ \frac{\mu}{\lambda e }\frac{\mathcal{L}'_{n-1}(n\nu)}{\mathcal{L}_{n-1}(n\nu)}
  \left(\ii,-\sqrt{1-e^2},0
 \right),
\end{equation}
 where $\nu$ is as defined in equation (\ref{NU}).

\begin{lemma}\label{Zlemma}
Define,
\[\bm{Z}_{0,\infty}:=\lim\limits_{\genfrac{}{}{0pt}{}{n\uparrow\infty, \epsilon\downarrow 0}{\lambda = n\epsilon^2}}
\bm{Z}_{\epsilon,n}.\]
Then,
\[
\vec{Z}_{0,\infty}(\vec{x})
= \frac{\ii\mu}{2\lambda }\left(
   1+\sqrt{1-\frac{4}{\nu}}\right)
    \frac{\vec{x}}{|\vec{x}|}+ \frac{\mu}{2\lambda e }\left(1-\sqrt{1-\frac{4}{\nu}}\right)
  \left(\ii,-\sqrt{1-e^2},0
 \right),
 \]
where $\vec{x} = (x,y,z)$  and $\nu$ is as defined in equation (\ref{NU}).
\end{lemma}
\begin{proof}
See \cite{Divine}
\end{proof}

\begin{lemma}\label{alltheRS}
Let,
\[\bm{Z}_{0,\infty} (\bm{x}) = \epsilon^2\nabla(S_{\epsilon}-iR_{\epsilon}).\]
Then,
\begin{eqnarray*}
\nabla R_{\epsilon} & = &  -\frac{\mu}{2e\lambda\epsilon^2 }\left\{
(1+\alpha)\frac{e\bm{x}}{|\bm{x}|}+\left((1-\alpha),\beta\sqrt{1-e^2},0\right)\right\}\\
\nabla S_{\epsilon} & = &-\frac{\mu}{2e\lambda\epsilon^2 }\left\{
\frac{\beta e\bm{x}}{|\bm{x}|}+\left(-\beta,(1-\alpha)\sqrt{1-e^2},0\right)\right\},
\end{eqnarray*}
 where
 \begin{eqnarray*}
 \alpha & = & \left(\hf\sqrt{\frac{\left(e|\vec{x}|-x-4\lambda^2 e/\mu\right)^2 +(1-e^2)y^2}{(e |\vec{x}|-x)^2+(1-e^2)y^2}}+\right.\\
 &  &\qquad\left.+\hf\frac{\left(e|\vec{x}|-x-2\lambda^2e/\mu\right)^2 +(1-e^2)y^2
 -4\lambda^4 e^2/\mu^{2}}{(e |\vec{x}|-x)^2+(1-e^2)y^2}\right)^{\hf}_,\\
 \beta & = & \frac{- 2\lambda^2 e \sqrt{1-e^2} y}{\mu\left((e|\vec{x}|-x)^2+(1-e^2)y^2\right)\alpha}.
 \end{eqnarray*}
\end{lemma}
\begin{proof} A simple calculation from Lemma \ref{Zlemma}.
\end{proof}

By setting,
\[\bm{b} = \lim\limits_{\genfrac{}{}{0pt}{}{n\uparrow\infty, \epsilon\downarrow 0}{\lambda = n\epsilon^2}}
\epsilon^2\nabla(R_{\epsilon,n}+S_{\epsilon,n})
 = \mathrm{Re}( \bm{Z}_{0,\infty})-\mathrm{Im}(\bm{Z}_{0,\infty}) = \epsilon^2\nabla(R_{\epsilon}+S_{\epsilon}),\]
 we find the following limiting Nelson diffusion:

\begin{theorem}\label{process}
The limiting Nelson diffusion  $\vec{X}_{\epsilon}$ satisfies the It\^{o} equation,
\[
 \di \vec{X}_{\epsilon}(s)  = \vec{b}(\vec{X}_{\epsilon}(s))\di s +\epsilon \di \vec{B}(s),
 \]
 where $\vec{b}(\vec{x}) = (b_x,b_y,b_z)$ in cartesian coordinates with,
 \begin{eqnarray*}
 b_x & = & \frac{\mu}{2\lambda}\left\{(\alpha+\beta-1)\frac{1}{e} - (\alpha+\beta+1)\frac{x}{|\vec{x}|}\right\}, \\
 b_y & = & \frac{\mu}{2\lambda}\left\{(\alpha-\beta-1)\frac{\sqrt{1-e^2}}{e} - (\alpha+\beta+1)\frac{y}{|\vec{x}|}\right\},\\
 b_z & = & -\frac{\mu}{2\lambda}(\alpha+\beta+1)\frac{z}{|\vec{x}|},
 \end{eqnarray*}
 with $\alpha$ and $\beta$ as defined in Lemma \ref{alltheRS}.
\end{theorem}
\begin{proof} See \cite{Divine}.
\end{proof}

 As a result of taking this limit, $\bm{X}_{\epsilon}$  does not satisfy the Nelson-Newton law and so is not a true Nelson diffusion. This is because it is not associated with an exact solution to the stationary Schr\"{o}dinger equation. However, if we let $\psi_{\epsilon} = \exp(R_{\epsilon}+i S_{\epsilon})$, we find an approximate wave function which is the correspondence limit of the atomic elliptic state.

\begin{figure}[h]
\centering
\resizebox{120mm}{!}{\includegraphics{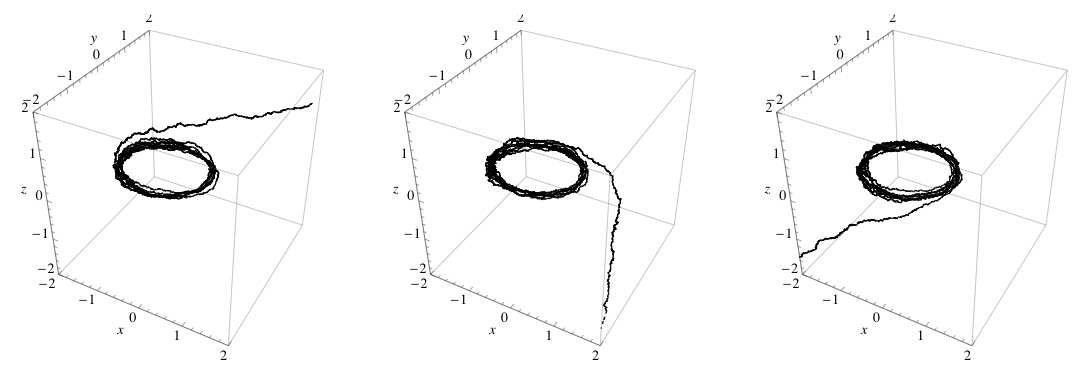}}
\caption{Simulations of the limiting Nelson diffusion $\bm{X}_{\epsilon}$ for $e=0.5$, $\epsilon=0.1$.}
\label{sims}
\end{figure}

\begin{theorem}\label{limitingwave}
The correspondence limit of the wave function $\psi_{\epsilon,n}$ is the limiting wave function $ \psi_{\epsilon}$, where
\[\psi_{\epsilon}(\vec{x}) = \nu^{\frac{\lambda}{\epsilon^2}}\left(1+\sqrt{1-\frac{4}{\nu}}\right)^{\frac{2\lambda}{\epsilon^2} }\exp\left(-\frac{\mu }{\lambda\epsilon^2}|\vec{x}| + \frac{\lambda \nu}{2\epsilon^2}\left(1-\sqrt{1-\frac{4}{\nu}}\right)\right),\]
and $\nu$ is as defined in equation (\ref{NU}).
\end{theorem}
\begin{proof} See \cite{Divine}.
\end{proof}
Simulations of  the limiting Nelson diffusion are shown in Figure \ref{sims}. The paths all rapidly converge to a neighbourhood of the Kepler ellipse.

\begin{remark}
As expected from equation (\ref{SBE1}), the vector $\bm{Z}_{0,\infty}$ satisfies,
\begin{equation}\label{ZEQN1}
\frac{1}{2} \bm{Z}_{0,\infty}^2 -\frac{\mu}{|\bm{x}|} = -\frac{\mu^2}{2\lambda^2}.
\end{equation}
Consequently, by taking imaginary parts of equation (\ref{ZEQN1}) we find,
\begin{equation}\label{orthog}
\nabla R_{\epsilon} \cdot \nabla S_{\epsilon} =0,
\end{equation}
which will be key in what follows.
\end{remark}

\subsection{The drift singularity}
\begin{figure}[h]
\centering
\resizebox{100mm}{!}{\includegraphics{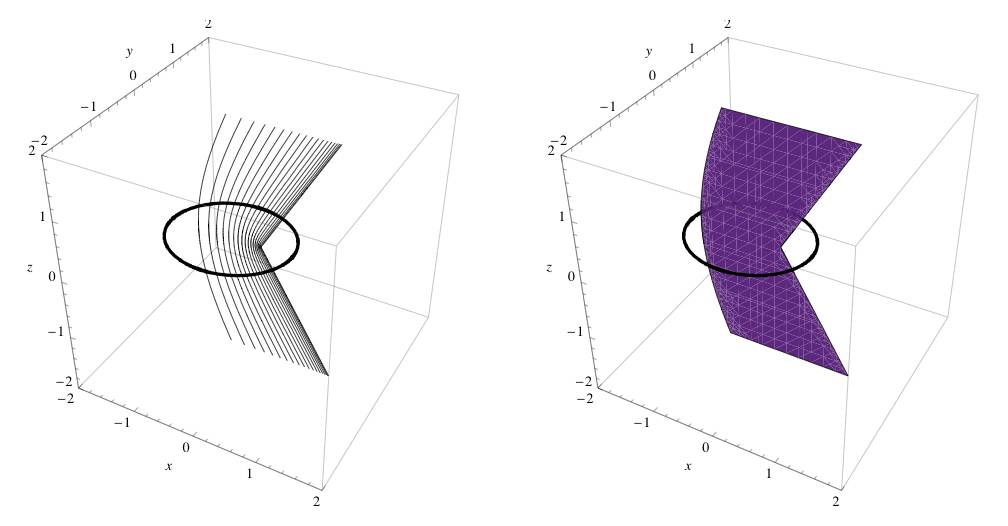}}
\caption{The nodal surfaces  for $\psi_{\epsilon,n}$ for $e=0.5$ and  $n=10$ together with their correspondence limit $\Sigma$ shown with the Kepler ellipse.}
\label{nodal}
\end{figure}

The limiting Nelson diffusion  $\bm{X}_{\epsilon}$ has a singularity in the drift field $\bm{b}$ which arises from the existence of nodes in the atomic elliptic state wave function $\psi_{\epsilon,n}$. The nodes of $\psi_{\epsilon,n}$ are given by the point  $|\bm{x}|=0$ and also by the right branch of a  series of $n-1$ hyperbolas in the plane $y=0$.
The drift $\bm{b}_{\epsilon,n}$ of the Nelson diffusion  $\bm{X}_{\epsilon,n}$  has infinite blowups on these curves. This behaviour makes the nodes inaccessible to the Nelson diffusion \cite{MR882809}.  The drift $\bm{b}$ for the limiting Nelson diffusion $\bm{X}_{\epsilon}$ also has an infinite blowup at the point $|\bm{x}|=0$. However, the correspondence limit of the hyperbolas forms a finite jump discontinuity in $\bm{b}$ across the surface $\alpha =0$. This  surface of discontinuity is,
\[\Sigma = \left\{(x,0,z):\, \frac{-e \left(4\lambda^2/\mu-\sqrt{\left(16 \lambda^4/\mu^2-z^2\right) e^2+z^2}\right)}{1-e^2}<x< \sqrt{\frac{e^2z^2}{1-e^2}}\right\}.
\]
This limiting singularity and the nodal curves for the exact case are shown in Figure \ref{nodal}. They are both ringed by the Kepler ellipse.

\subsection{Unifying the two and three dimensional analyses}
We will now prove that our results in \cite{Divine} for the two dimensional case are relevant for the full three dimensional problem.
Recall that for integer $n$ the three dimensional atomic elliptic state $\psi_{\epsilon,n}(\bm{x},t)$ satisfies the stationary Schr\"{o}dinger equation,
\[-\hf\epsilon^4\Delta\psi_{\epsilon,n}(\vec{x})-\frac{\mu}{|\vec{x}|}\psi_{\epsilon,n}(\vec{x}) = E_n\psi_{\epsilon,n}(\vec{x}),\]
where $\epsilon^2 = \hbar$, $E_n = -\mu^2/(2\lambda^2)$, $\lambda  = n\epsilon^2$ and  $\vec{x} = (x,y,z)$.

It can be shown that the two dimensional function,
\[
\psi_{\epsilon,n}^{2d}(\vec{x})= \exp\left(-\frac{(2n+1)\mu }{2\lambda^2}|\vec{x}|\right) \mathcal{H}_{2n}\left(\sqrt{2n+1}\sqrt{\hf[\nu]}\right),
\]
where $\bm{x} = (x,y)$, $\nu$ is as in equation (\ref{NU}) and $\mathcal{H}_{2n}$ is a Hermite polynomial, satisfies the stationary Schr\"{o}dinger equation,
\[-\hf\epsilon^4\Delta\psi_{\epsilon,n}^{2d}(\vec{x})-\frac{\mu}{|\vec{x}|}\psi_{\epsilon,n}^{2d}(\vec{x}) = E_n\psi_{\epsilon,n}^{2d}(\vec{x}),\]
where again $\epsilon^2 = \hbar$, $E_n = -\mu^2/(2\lambda^2)$,  but now $\lambda  = n\epsilon^2+\epsilon^2/2$.

We will investigate the connection between the correspondence limits of these two and three dimensional wave functions. For this we need the following result on Hermite polynomials. We follow the conventions in \cite{MR0240343}.

\begin{lemma}\label{Qlemma}
Let,
\[Q_m =\frac{1}{\sqrt{m+1} }\times \frac{\mathcal{H}_m'\left(\sqrt{m+1}\sqrt{\nu/2}\right)}{\mathcal{H}_m\left(\sqrt{m+1}\sqrt{\nu/2}\right)},\]
where $\mathcal{H}_m$ denotes the $m^{\mathrm{th}}$ Hermite polynomial. Then,
\[\lim\limits_{\genfrac{}{}{0pt}{}{n\uparrow\infty, \epsilon\downarrow 0}{\lambda =
 n\epsilon^2+\epsilon^2/2}}\frac{Q_{2n}}{\sqrt{\nu/2}} = 1-\sqrt{1-\frac{4}{\nu}}.\]
\end{lemma}
\begin{proof}
Clearly,
\[Q_m
= \frac{2m}{\sqrt{m+1} }\times \frac{\mathcal{H}_{m-1}\left(\sqrt{m+1}\sqrt{\nu/2}\right)}{ \mathcal{H}_m\left(\sqrt{m+1}\sqrt{\nu/2}\right)}. \]
Then using the standard recurrence relation,
\[\mathcal{H}_{m+1}(v) = 2v\mathcal{H}_m(v)-2 m \mathcal{H}_{m-1}(v),\]
with $v=\sqrt{m+1}\sqrt{\nu/2}$ we can conclude,
\[\frac{2\sqrt{m+1}}{\sqrt{m+2}}\times\frac{1}{Q_{m+1}} = 2 \sqrt{\nu/2}-Q_m.\]
Assuming that $Q_m \rightarrow q$ as $m\rightarrow\infty$ then,
\[\frac{2}{q} = 2\sqrt{\nu/2}-q,\]
giving the desired result.
\end{proof}

\begin{theorem}\label{2d3d}Let $\psi_{\epsilon}^{3d}(x,y,z)$ denote the correspondence limit (in the sense of Theorem \ref{limitingwave}) of the three dimensional atomic elliptic state $\psi_{\epsilon,n}(x,y,z)$ and let $\psi_{\epsilon}^{2d}(x,y)$ denote the correspondence limit of the two dimensional wave function $\psi_{\epsilon,n}^{2d}(x,y)$. Then,
\[\psi_{\epsilon}^{2d}(x,y) = \psi_{\epsilon}^{3d}(x,y,z)|_{z=0}.\]
\end{theorem}

\begin{proof}
If we define,
\[\bm{Z}_{\epsilon,n}^{2d}(\bm{x}) = -i\epsilon^2\nabla\ln\psi_{\epsilon,n}^{2d},\]
then in Cartesians,
\[\bm{Z}_{\epsilon,n}^{2d}(\bm{x}) =  \frac{i\mu}{\lambda} \left(1-\frac{Q_{2n}}{2\sqrt{\nu/2}}\right)\frac{ \bm{x}}{|\bm{x}|}
+\frac{\mu}{\lambda e}\,\frac{Q_{2n}}{2\sqrt{\nu/2}}\left(i,-\sqrt{1-e^2}\right),\]
where $Q_m$ is as defined in Lemma \ref{Qlemma}. Taking the correspondence limit
gives
\begin{eqnarray*}
 \bm{Z}_{0,\infty}^{2d}&:=&\lim\limits_{\genfrac{}{}{0pt}{}{n\uparrow\infty, \epsilon\downarrow 0}{\lambda = n\epsilon^2+\epsilon^2/2}}
\bm{Z}^{2d}_{\epsilon,n}  \\
& = &
\frac{i\mu}{2\lambda} \left(1+\sqrt{1-\frac{4}{\nu}}\right)\frac{ \bm{x}}{|\bm{x}|}
+\frac{\mu}{2\lambda e}\left(1-\sqrt{1-\frac{4}{\nu}}\right)\left(i,-\sqrt{1-e^2}\right),
\end{eqnarray*}
where $\bm{x} = (x,y)$. Therefore, from Lemma \ref{Zlemma} we can conclude,
\[(\bm{Z}_{0,\infty}^{2d}(x,y),0) \equiv \bm{Z}_{0,\infty}(x,y,z)|_{z=0},\]
which gives the result.
\end{proof}

In \cite{Divine} we worked with the limiting Nelson diffusion restricted to the plane $z=0$ which we can now view as being derived from the correspondence limit of a  two dimensional wave function. We showed that as $\epsilon\rightarrow 0$ this two dimensional limiting Nelson diffusion converged to the deterministic dynamical system $\bm{X}_0$, where
\[\dot{\bm{X}}_0 = \bm{b}(\bm{X}_0),\]
provided the path avoided the drift singularity $\Sigma$. Moreover, we showed that the Kepler ellipse was an asymptotically stable periodic orbit for this dynamical system whose domain of attraction contained all of $\mathbb{R}^2$ outside the Kepler ellipse. We also showed that a particle on the Kepler ellipse would move according to Kepler's laws of planetary motion. Thus, we could conclude that in the infinite time limit, any trajectory beginning outside the Kepler ellipse would converge to Keplerian motion on the Kepler ellipse. These results were proved using the following non-orthogonal coordinate system which we will again use:
\begin{defn}
The Keplerian elliptic coordinates $(u,v)$ in the plane $(x,y)$ are defined by,
\[
 x = \frac{2 a e (\cos v - u)}{e+u},\quad  y = \frac{2 a e \sqrt{1-u^2} \sin v}{e+u},
  \]
where $-e<  u \le 1$ and $0\le v<2 \pi $ with $a = \lambda^2/\mu$.
\end{defn}
The $u$ coordinate curves form a family of non-confocal ellipses with eccentricity $|u|$, foci at $(0,0)$ and $(-4 a e u/(1+e),0)$, and semimajor axis $2 a e/(e+u)$ parallel to the $x$ axis. The Kepler ellipse is given by the coordinate curve $u=e$, the singularity $\Sigma$ by the curve $u=1$ and the ellipse at infinity by $u=-e$. On the Kepler ellipse $v$ is the eccentric angle.

Unfortunately we have been unable to find an extension for this system to simplify the three dimensional limiting Nelson diffusion. However, the following naive extension will be of use in studying the invariant measure.
 \begin{defn}
The cylindrical  Keplerian elliptic coordinates $(u,v,z)$ in three space are defined by,
\[
 x = \frac{2 a e (\cos v - u)}{e+u},\quad  y = \frac{2 a e \sqrt{1-u^2} \sin v}{e+u}, \quad z=z, 
  \]
where $-e<  u \le 1$, $0\le v<2 \pi $, $z\in \mathbb{R}$ with $a = \lambda^2/\mu$.
\end{defn}

   It follows from Theorem \ref{2d3d} that if we could prove that the motion was confined to a neighbourhood of the plane $z=0$ then our two dimensional results would be relevant. This provides the motivation for studying the invariant measure for the full three dimensional problem.
\section{The invariant measure for $\bm{X}_{\epsilon}$}

We again recall that the atomic elliptic state $\psi_{\epsilon,n} = \exp(R_{\epsilon,n} + i S_{\epsilon,n})$ satisfies  the stationary Schr\"{o}dinger equation (\ref{SSE2}) provided $n$ is an integer. Therefore, assuming $n$ is an integer, the Nelson diffusion  $\bm{X}_{\epsilon,n}$ satisfies the Nelson-Newton law and has invariant density $\rho_{\epsilon,n} = \exp(2 R_{\epsilon,n})$. Thus the forward Kolmogorov equation (\ref{FK1}) gives,
\[0 = \Delta S_{\epsilon,n} + 2 \nabla R_{\epsilon,n} \cdot \nabla S_{\epsilon,n}.\]
Moreover,  from equation (\ref{AES1}) it is possible to write,
\[R_{\epsilon,n} = \tilde{R}_{\epsilon,n}/\epsilon^2,\quad S_{\epsilon,n} = \tilde{S}_{\epsilon,n}/\epsilon^2,\]
where $\tilde{R}_{\epsilon,n}$ and $\tilde{S}_{\epsilon,n}$ only depend on $\epsilon$ through $\lambda = n\epsilon^2$. Thus,
\begin{equation}\label{CE3}0 = \epsilon^2\Delta \tilde{S}_{\epsilon,n} + 2 \nabla\tilde{R}_{\epsilon,n} \cdot \nabla \tilde{S}_{\epsilon,n}.
\end{equation}
This equation is equivalent to $\rho_{\epsilon,n}$ being the density for the invariant measure for $\bm{X}_{\epsilon,n}$.

The limiting Nelson diffusion  $\bm{X}_{\epsilon}$ does not satisfy the Nelson-Newton law and the limiting state $\psi_{\epsilon}=\exp(R_{\epsilon}+iS_{\epsilon})$ is only an approximate wave function. Therefore we cannot automatically conclude that $\exp(2 R_{\epsilon})$ is the invariant density for $\bm{X}_{\epsilon}$. However, we can write $R_{\epsilon} = R/\epsilon^2$ and $S_{\epsilon} = S/\epsilon^2$ and then from equation (\ref{CE3}) conclude that $\exp(2 R_{\epsilon})$ would be the invariant measure if and only if,
\begin{equation}\label{CE4}0 = \epsilon^2\Delta S + 2 \nabla R \cdot \nabla S.
\end{equation}
Thus, from equation (\ref{orthog}) with the observation $\Delta S\neq 0$ we find that  $\exp(2 R_{\epsilon})$ is the invariant measure for the limiting Nelson diffusion only when $\epsilon\rightarrow 0$.
We now show that this is the physically correct invariant measure.
\begin{lemma}
Let the three dimensional limiting wave function be given by $\psi_{\epsilon} = \exp\left(R_{\epsilon}+iS_{\epsilon}\right)$ where $R_{\epsilon}$ and $S_{\epsilon}$ are real. Then,
\[\frac{\partial R_\epsilon}{\partial z}(x,y,z) = 0 \quad\Leftrightarrow \quad z=0.\]
\end{lemma}
\begin{proof}
 Follows from Lemma \ref{alltheRS}.
\end{proof}

\begin{theorem}
$R_{\epsilon}(x,y,z)$ has a manifold of constant maxima on the Kepler ellipse in the plane $z=0$.
\end{theorem}
\begin{proof}
From the previous lemma we know that $z=0$ is a necessary condition for $R_{\epsilon}$ to have a critical point
and so we can write $\alpha$ and $\beta$ from Lemma  \ref{alltheRS} in Keplerian elliptic coordinates as,
\[\alpha = \frac{\sqrt{(1-u^2)(1-e^2)}}{1+eu-(e+u)\cos v},\qquad \beta = -\frac{(e+u)\sin v}{1+ eu-(e+u)\cos v}.
\]
The Kepler ellipse is the curve $u=e$ and so again from Lemma \ref{alltheRS} we find that $\left.\nabla R_{\epsilon}\right|_{z=0, u=e}= 0$. Moreover, it can easily be shown that on the Kepler ellipse,
\[\left.R_{\epsilon}\right|_{z=0, u=e} =\frac{\lambda}{2 \epsilon^2}\ln\left(\frac{16}{e^2}\right) ,\]
so that $R_{\epsilon}$ is constant around the ellipse. Clearly, any point on the Kepler ellipse is therefore a degenerate critical point of $R_{\epsilon}$ forcing the Hessian of $R_{\epsilon}$ to be singular.
However, in cylindrical Keplerian elliptic coordinates $(u,v,z)$ the Hessian on the Kepler ellipse is,
\[\left.R_{\epsilon}''(u,v,z)\right|_{z=0, u=e} = \left(
\begin{array}{ccc}-
 \frac{\lambda \left(1+e^2+2 e \cos v\right) }{4 e^2 \epsilon ^2 (1-e^2)} & 0 & 0 \\
 0 & 0 & 0 \\
 0 & 0 & -\frac{\mu ^2}{\epsilon ^2 \lambda ^3 \left(1+e^2-2 e\cos v\right)}
\end{array}
\right)\]
confining the degeneracy to the $v$ coordinate as expected. It is now clear that this must be a manifold of maxima.
\end{proof}
This maximum forms a volcano shape in the plane $z=0$ for $\exp(2R_{\epsilon})$ as is shown in Figure \ref{maximum}.
By considering the asymptotic behaviour of this function as $\epsilon\sim 0$ we can find the thickness of this volcano in the $z=0$ plane and also the thickness in the $z$ direction:
\begin{figure}[h]
\centering
\resizebox{130mm}{!}{\includegraphics{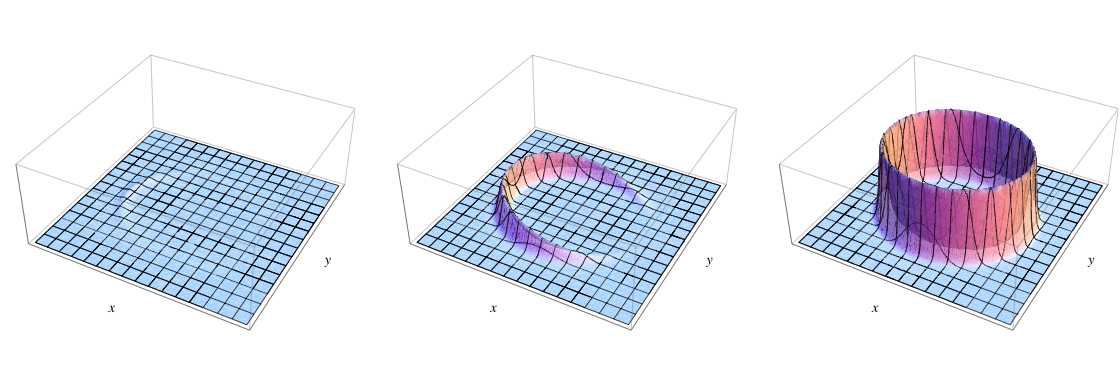}}
\caption{ $\exp(2R_{\epsilon})$ for $z=0.2$, $z=0.1$ and $z=0$ with $e = 0.5$ and $\epsilon=0.1$.}
\label{maximum}
\end{figure}
\begin{theorem}\label{laplace}
 For an arbitrary test function $G\in\mathscr{S}(\mathbb{R}^3)$,
\[\iiint_{\mathbb{R}^3}
G(\bm{x})\exp\left(\frac{2 R (\bm{x})}{\epsilon^2}\right)\di \bm{x} \sim
 \frac{\pi a^3   \epsilon ^2 }{\lambda} \left(\frac{16}{e^2}\right)^{\frac{\lambda }{\epsilon ^2}}
\int_0^{2\pi} G(v) g(v) \di v
\]
as $\epsilon\sim 0$,
where
\[g(v)=(1-e \cos v) \sqrt{1+e^4-2 e^2\cos 2 v},\]
\[G(v) = G(\bm{x}(v)),\quad \bm{x}(v) = (a\cos v-ae,a\sqrt{1-e^2}\sin v,0),\]  and
$v$ is the eccentric angle of $\bm{x}(v)\in\mathcal{E}_e$.
\end{theorem}
\begin{proof}
We can write the integral using cylindrical Keplerian elliptic coordinates and then apply Laplace's method twice,
\begin{eqnarray*}
\lefteqn{\iiint_{\mathbb{R}^3}
G(\bm{x})\exp\left(\frac{2 R (\bm{x})}{\epsilon^2}\right)\di \bm{x}}\\
& = &
\int_0^{2\pi}\int_{-e}^1\int_{-\infty}^{\infty}
G(u,v,z)\exp\left(\frac{2 R (u,v,z)}{\epsilon^2}\right)
\left|\frac{\partial(x,y,z)}{\partial (u,v,z)}\right|\di z\di u\di v\\
& \sim &
\int_0^{2\pi}G(e,v,0)
\exp\left(\frac{2 R (e,v,0)}{\epsilon^2}\right)\left[\left|\frac{\partial(x,y,z)}{\partial (u,v,z)}\right|\sqrt{\frac{\pi \epsilon^2}{\left|\frac{\partial^2 R}{\partial z^2}\right|}}
\sqrt{\frac{\pi \epsilon^2}{\left|\frac{\partial^2 R}{\partial u^2}\right|}}\right]_{z=0, u=e}
\di v.
\end{eqnarray*}
The result then follows from a simple calculation.
\end{proof}
\begin{cor}\label{ellipticE} As $\epsilon\sim 0$,
\[
\iiint_{\mathbb{R}^3}\exp\left(\frac{2 R (\bm{x})}{\epsilon^2}\right)\!\di \bm{x} \sim
 \left(\frac{16}{e^2}\right)^{\frac{\lambda }{\epsilon ^2}}\frac{2  \pi a^3  \epsilon ^2 }{\lambda }
\bigg\{(1-e^2) E(-\xi_-^2)+(1+e^2) E(\xi_+^2)\bigg\},
\]
where $E(\xi)$ denotes the complete elliptic integral of the second kind and 
\[\xi_{\pm} = \frac{2 e}{1\pm e^2}.\]
 \end{cor}
 \begin{proof} Follows directly from Theorem \ref{laplace}.
 \end{proof}

\begin{cor}
The effective standard deviation of the Gauss measure,
\[\frac{\exp\left(\frac{2 R(\bm{x})}{\epsilon^2}\right)}{\iiint_{\mathbb{R}^3}
\exp\left(\frac{2 R (\bm{x})}{\epsilon^2}\right)\di \bm{x} },\]
 measured normal to the Kepler ellipse in the plane $z=0$ is, to leading order in $\epsilon$,
\[\frac{\epsilon \lambda^{\hf[3]}}{\mu}\sqrt{\frac{ (1-e \cos v) \left(1+e^2+2e \cos v\right)}{1+e \cos v}},\]
and in the $z$ direction is,
\[\frac{\epsilon \lambda^{\hf[3]}}{\mu}\sqrt{1+e^2-2e \cos v }.\]
\end{cor}
\begin{proof}
Combining Corollary \ref{ellipticE} with Taylor's formula gives,
\begin{equation}\label{expansion}
\frac{\exp\left(\frac{2 R(\bm{x}_e+\epsilon \bm{h})}{\epsilon^2}\right)}{\iiint_{\mathbb{R}^3}
\exp\left(\frac{2 R (\bm{x})}{\epsilon^2}\right)\di \bm{x} }
 \sim
 \frac{k}{\epsilon^2} \exp\left(\sum_{i,j} \frac{\partial^2 R}{\partial x_i\partial x_j}(\bm{x}_e)h_i h_j+O(\epsilon)\right)
 \end{equation}
 for some constant $k$ where $\bm{x}_e$ is a point on the Kepler ellipse and $\bm{h}=(h_1,h_2,h_3)$ is a vector in the plane through $\bm{x}_e$ orthogonal to the Kepler ellipse.
Since $R$ has a maximum on the Kepler ellipse, its Hessian is negative definite and so the leading term in equation  (\ref{expansion}) can be written as,
 \[\exp\left(-|\bm{h}|^2\left|\sum_{i,j} \frac{\partial^2 R}{\partial x_i\partial x_j}(\bm{x}_e)\hat{h}_i \hat{h}_j\right|\right),\]
where $\hat{\bm{h}}$ is the unit vector parallel to $\bm{h}$. 
Therefore, we can conclude that the width of the volcano in the direction $\hat{\bm{h}}$ can be measured in units,
\[\epsilon \left|\sum_{i,j} \frac{\partial^2 R}{\partial x_i\partial x_j}(\bm{x}_e)\hat{h}_i \hat{h}_j\right|^{-\hf},\]
giving the result.
 \end{proof}

We now look to find an invariant density for $\bm{X}_{\epsilon}$ when $\epsilon\neq 0$. We follow the methods of Truman and Zhao \cite{MR1628310} (for zero noise) and make the ansatz that the invariant density can be written in the form,
\begin{equation}\label{density}
\rho_{\epsilon}(\bm{x}): = T(\bm{x}) \exp \left(\frac{2{R}(\bm{x})}{\epsilon^2}\right) = \exp\left( \frac{2{R}(\bm{x})}{\epsilon^2}+\ln T(\bm{x})\right),
\end{equation}
where $T$ is $O(1)$ in $\epsilon$. As $\epsilon\sim 0$, the exponential term will vary  rapidly in space variables, with a pronounced peak of constant maxima on the Kepler ellipse in the plane $z=0$, while $T$ will only vary slowly in space variables,  this variation being in the direction tangential to the Kepler ellipse. Let $\pi_{\epsilon}$ denote the stationary probability measure with density (before normalisation) given by $\rho_{\epsilon}$. We can show that this invariant measure is concentrated on the Kepler ellipse for small $\epsilon$:
\begin{cor} \label{cor2}
For an arbitrary test function $f\in\mathscr{S}(\mathbb{R}^3)$, if $\mathbb{E}$ denotes expectation with respect to $\pi_{\epsilon}$,
\[\mathbb{E}\left\{f\right\}\sim
\frac{\int_0^{2\pi} f(v) T(v) g(v)\di v}
{\int_0^{2\pi} T(v)g(v)\di v}
\]
as $\epsilon\sim 0$, where $f(v) = f(\bm{x}(v))$ with $\bm{x}(v)\in\mathcal{E}_e$ and $g(v)$ as in Theorem \ref{laplace}.
\end{cor}
\begin{proof}
Follows from Theorem \ref{laplace}.
\end{proof}

\begin{figure}[h]
\centering
\resizebox{100mm}{!}{\includegraphics{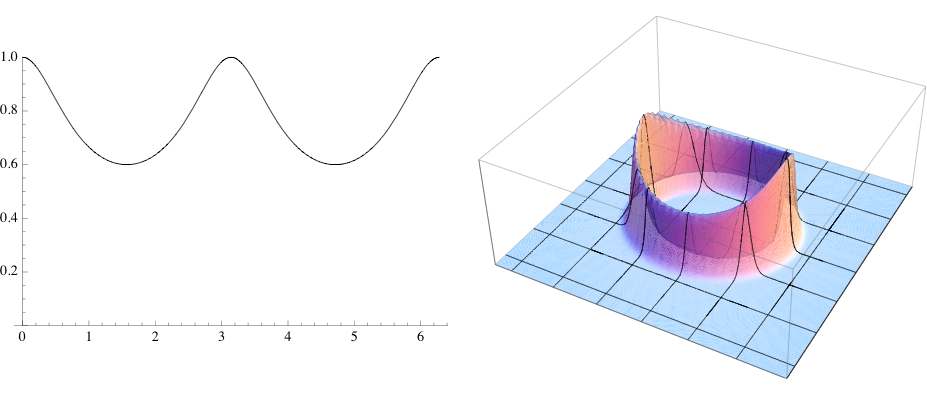}}
\caption{The function $T(v)$ with the invariant density $\rho_{\epsilon}$.}
\label{Tmeasure}
\end{figure}

We can also find the function $T$ up to leading order in $\epsilon^2$.
Assuming $\rho_{\epsilon}$ is the invariant density, the forward Kolmogorov equation (\ref{FK1})  yields,
\[\nabla\cdot\left(\left(-\nabla S + \hf[\epsilon^2]\nabla \ln T\right)\rho_{\epsilon}\right)=0,\]
giving,
\[-\nabla\ln T\cdot\left(\nabla S-\nabla R -\epsilon^2\nabla\ln T\right)
-\Delta S +\hf[\epsilon^2]\Delta \ln T=0,\]
which from equations (\ref{BD1}) and (\ref{OV1}) can be written,
\[\nabla\ln T\cdot \bm{b}_- +\nabla\cdot\bm{v} =0.\]
If we take only the leading term in $\epsilon^2$ then,
\[(\nabla\ln T)\cdot\nabla(S-R) + \Delta S = 0.\]
  Assuming the variation in $T$ is parallel to the Kepler ellipse we can write $\nabla \ln T = \gamma\bm{b}$ as the drift $\bm{b}$ is tangential to the Kepler ellipse when the particle is on the ellipse. Moreover, on the Kepler ellipse $\nabla R=0$ and so,
\[\nabla \ln T = -\Delta S\frac{ \nabla S}{|\nabla S|^2}.\]
Using the Keplerian elliptic coordinates on the Kepler ellipse so that $v$ becomes the eccentric angle,  we see that the variation in $T$ as the particle precesses around the Kepler ellipse is,
\[\frac{\di}{\di v}\ln T =\nabla\ln T \cdot \frac{\di \bm{x}(v)}{\di v} = -\frac{2 e^2 \sin 2 v}{1+e^4-2e^2 \cos 2 v}. \]
Hence,
\[T(v) = \exp\left(-\int_0^{v}\frac{2 e^2 \sin 2 \tilde{v}}{1+e^4-2 e^2\cos 2 \tilde{v}}\di\tilde{v}\right) = \frac{(1 - e^2)}{\sqrt{1 + e^4 - 2 e^2 \cos 2v}}.\]
This gives the slowly varying multiplicative behaviour of the invariant density $\rho_{\epsilon}$ which is shown in Figure \ref{Tmeasure}. This  behaves exactly as one would expect from the results in \cite{Lena}.

\begin{cor}
For an arbitrary test function $f\in\mathscr{S}(\mathbb{R}^3)$, if $\mathbb{E}$ denotes expectation with respect to $\pi_{\epsilon}$,
\[\mathbb{E}\left\{f\right\}\sim
\frac{1}{2 \pi}\int_0^{2\pi}f(v) (1 - e \cos v)\di v,
\]
as $\epsilon\sim 0$, where $f(v) = f(\bm{x}(v))$ and $\bm{x}(v)\in\mathcal{E}_e$.
\end{cor}
\begin{proof}
Follows from Corollary \ref{cor2} with,
\[T(v) g(v)  = (1-e^2)(1-e\cos v).\]
\end{proof}

\section{Some preliminary results}
\begin{figure}
\center
\setlength{\unitlength}{6mm}
\begin{picture}(12,8)
\put(0,0){\resizebox{72mm}{!}{\includegraphics{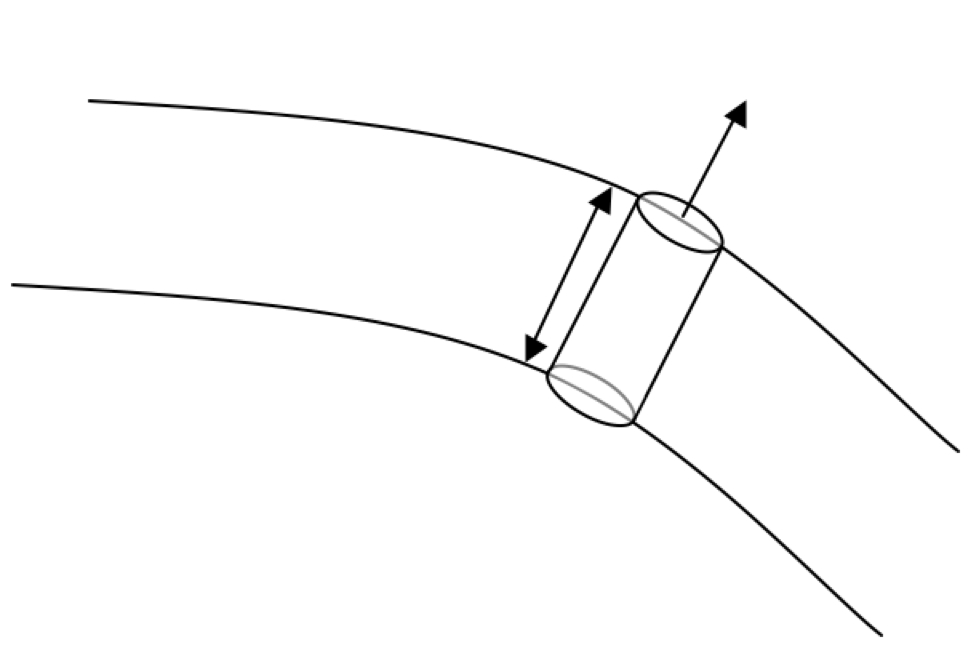}}}
\put(6.5,4.5){$\sigma$}
\put(8.9,7){$\hat{\nabla f}$}
\put(6.7,2.4){$\di\mu_{\partial}$}
\put(2,3.6){$f^2=s$}
\put(2,6.9){$f^2=s+\di s$}
\end{picture}
\caption{The elementary co-area formula}
\label{coareapic}
\end{figure}
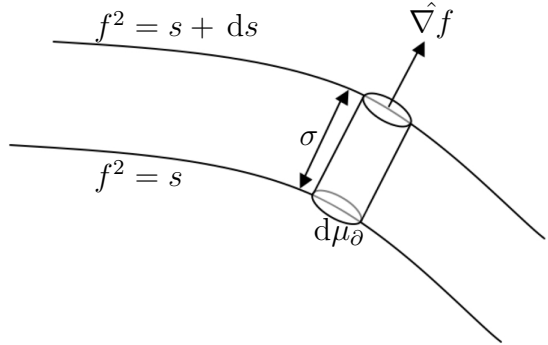

Before we discuss the spectral gap for the generator of the limiting Nelson diffusion process, we will first introduce some of the tools for our proof; the co-area formula, Cheeger's inequality and the Poincare inequality. The first two of these are standard results in Riemannian geometry which we will recast in Euclidean space. We let $M\subset \mathbb{R}^3$  be an open Borel set and let $\mu$ be a probability measure on the Borel sets of $M$ with a smooth density $\rho$ with respect to $3$-dimensional Lebesgue measure. Let $\mu_{\partial}$ be the $2$-dimensional measure with density $\rho$ with respect to $2$-dimensional Riemannian volume on surfaces.
\begin{lemma}[An elementary co-area formula]\label{coarea}
Let $f:M\rightarrow\mathbb{R}$ be a smooth measurable function. Then , if  $m := \sup_{x\in M} f^2(x)$,
\[\int_0^{m} \mu_{\partial}(f^2=s)\di s = \int_M |\nabla f^2|\di \mu.\]
\end{lemma}
\begin{proof}
Consider the level surfaces of the function $f^2$ (see Figure \ref{coareapic}).
Let  $\hat{\bm{n}} = \nabla f/|\nabla f|$ denote the unit vector orthogonal to the surface $f^2=s$. Then for small $\sigma>0$,
\[f^2(x+\sigma\hat{\bm{n}})=s+\di s,\]
and so to first order,
\[\nabla(f^2)\cdot\sigma\hat{\bm{n}} =\di s.\]
Evidently $\di \mu = \sigma\di\mu_{\partial}$ giving,
\[\di \mu_{\partial}\di s = 2\di\mu_{\partial}\sigma|f||\nabla f| = 2\di\mu |f||\nabla f|,\]
and the result follows.
\end{proof}

The following inequality due to Cheeger \cite{cheeger} originates in the study of lower bounds for the eigenvalues of the Laplacian on a compact Riemannian manifold with Dirichlet boundary conditions.  Our sketch proof is based on the presentations of Buser and Chavel \cite{buser, chavel}.
\begin{theorem}[Cheeger's inequality]\label{cheeger}Let,
\[h = \inf\limits_{A} \frac{\mu_{\partial}(\partial A)}{\mu(A)},\]
where $A\subset M$ varies over all open sets with smooth boundary satisfying $A\cap\partial M = \emptyset$, and let $f:M\rightarrow \mathbb{R}$ be a smooth measurable function with a suitable boundary condition e.g $f|_{\partial M} = 0$.
Then,
\[ \frac{ \left(\int |\nabla f|^2 \di \mu \right)}{\left(\int f^2 \di \mu\right)} \ge \frac{1}{4} h^2.\]
\end{theorem}

\begin{proof}
Let
$m := \max_{x\in M} f^2(x)$, then by the co-area formula (Lemma \ref{coarea}),
\[\int_M |\nabla f^2| \di\mu = \int_0^m  \mu_{\partial}(f^2=s)\di s \ge \tilde{h} \int_0^m  \mu(f^2\ge s)\di s,\]
where
\[\tilde{h} : = \inf\limits_{0\le s\le m} \frac{\mu_{\partial}(f^2=s)}{\mu(f^2\ge s)}.\]

Integration by parts gives,
\[
\int_0^m \mu(f^2\ge s)\di s =  - \int_0^m s\, \frac{\di}{\di s}\left( \mu(f^2\ge s)\right)\di s,\]
and,
\[\frac{\di}{\di s}\left( \mu(f^2\ge s)\right)
 = -\mu_{\partial}(f^2=s).\]
 Thus,
 \begin{eqnarray*}
 \int_M|\nabla f^2|\di \mu \ge \tilde{h} \int_0^m \mu(f^2\ge s) \di s &  = & \tilde{h} \int_0^m s\, \mu_{\partial} (f^2 = s) \di s\\
& = & \tilde{h} \int_M f^2 \di \mu_{\partial}.
\end{eqnarray*}
The Cauchy-Schwartz inequality gives,
\[\left(\int f^2 \di \mu\right) \left(\int |\nabla f|^2 \di \mu \right)\ge \left(\int f |\nabla f| \di \mu\right)^2,\]
and so,
\[\frac{ \left(\int |\nabla f|^2 \di \mu \right)}{\left(\int f^2 \di \mu\right)}
\ge \frac{\left(\int f |\nabla f| \di \mu\right)^2}{\left(\int f^2 \di \mu\right)^2} = \frac{\left(\int  |\nabla f^2| \di \mu\right)^2}{4\left(\int f^2 \di \mu\right)^2}\ge \frac{\tilde{h}^2}{4}.\]
Finally, it is clear that $\tilde{h}\ge h$ giving the result.\end{proof}

Our proof also uses the celebrated Poincar\'{e} inequality and its perturbation theory.
In particular we use the following classical Poincar\'{e} inequality for the Lebesgue measure of open balls in $\mathbb{R}^3$.
\begin{theorem} \label{poincare}Let $\di \bm{x}$ denote integration with respect to the $3$-dimensional Lebesgue measure. For all $f\in C^1(\mathbb{R}^d)$,
\[
 \int_{B_{r}} f^2 \di \bm{x} \le \frac{4 r ^2}{\pi^2} \int_{B_{r}}|\nabla f|^2\di \bm{x}+ \frac{1}{|B_{r }|} \left(\int_{B_{r}}f\di \bm{x}\right)^2,\]
where $|B_r|$ denotes the Lebesgue measure of the ball $B_r$.

\end{theorem}
\begin{proof}
A proof of the Poincar\'{e} inequality on Euclidean balls can be found in \cite{MR1872526}. The constant $4/\pi^2$ used in the inequality is derived from 
Theorem 1.3 of \cite{W94}. 
\end{proof}

We are now ready to prove the existence of the spectral gap for the generator of the limiting Nelson diffusion.
\section{The spectral gap}
Consider any Nelson diffusion process $\bm{X}$ with drift $\bm{b}$. The generator of this process is,
\[\mathcal{G} = \hf[\epsilon^2]\Delta + \bm{b}\cdot\nabla.\]
We can rewrite this operator using a simple similarity transformation.
\begin{lemma}\label{similarity}
For the generator $\mathcal{G}$ of any Nelson diffusion process,
\[\mathcal{G} = -\frac{1}{\epsilon^2}\exp(-R-S) \tilde{H} \exp(R+S),\]
where $\tilde{H}$ is the formal Hamiltonian,
\[\tilde{H} = \hf\left(-\epsilon^4 \Delta +  \epsilon^2 \nabla\cdot \bm{b} +\bm{b}^2\right).\]
Moreover for $\tilde{\psi} = \exp (R-S)$,
\[\tilde{H}\tilde{\psi} =0.\]
\end{lemma}
\begin{proof}
Simply follows from repeated application of the identity,
\[\Delta(fg) = f\Delta g + g\Delta f + 2 \nabla f\cdot\nabla g,\]
together with the condition $\bm{b} = \epsilon^2\nabla(R+S)$.
\end{proof}
Unfortunately, we have not been able to prove for the generator $\mathcal{G}$ of the limiting Nelson diffusion $\bm{X}_{\epsilon}$ defined by,
\[\mathcal{G} = \hf[\epsilon^2]\Delta + \bm{b}\cdot\nabla,\]
where $\bm{b}$ is as in Theorem \ref{process},
 that $\tilde{H}$ has a spectral gap as is explained in \cite{Divine}. Therefore, we have to resort to different techniques to understand the spectrum of this generator. Our aim here is to establish a suitable Poincar\'{e} inequality which will give us the existence of the spectral gap. This inequality is derived from separate inequalities which are established both inside and outside some open ball following a decomposition argument as in \cite{W99}. A detailed discussion of the use of Poincar\'{e} inequalities with diffusions can be found in \cite{Wang}.

For any $r>0$, define the open ball $B_{r}= \{\bm{x}\in\mathbb{R}^3: |\bm{x}|<r\}$ and its complement $B_{r}^c= \{\bm{x}\in\mathbb{R}^3: |\bm{x}|>r\}$. Let $\mathscr{L}_0(B_{r}^c)$ denote the set of all Lipschitz continuous functions with support in $B_{r}^c$. We assume that the invariant density for the limiting diffusion $\bm{X}_{\epsilon}$ can be written as,
\[\rho_{\epsilon} = T(\bm{x})\exp(2R_{\epsilon}(\bm{x}))\]
 and that there exists a constant $r_0>0$ such that,
\begin{equation}\label{assumption}|\nabla \ln T(\bm{x})|<C\end{equation}
for all $\bm{x}$ with $|\bm{x}|>r_0$ where   $\epsilon\in(0,1)$ is sufficiently small so that,
\[0<C<\frac{\mu}{\epsilon^2\lambda}.\]
We also assume that $\ln T(\bm{x})$ is bounded for all $\bm{x}\in\mathbb{R}^3$. If we define $V_{\epsilon} = 2 R_{\epsilon} + \ln T$ 
then we have $\di\pi_{\epsilon} = \exp(V_{\epsilon})\di\bm{x}$.

 Define a new operator,
\[\mathcal{G}_{\bm{u}}:= \hf[\epsilon^2]\Delta +(\bm{u}\cdot\nabla),\]
where $\bm{u} = (\epsilon^2/2) \nabla\ln\rho_{\epsilon}$ is the osmotic velocity in the stationary state corresponding to the invariant measure with density $\rho_{\epsilon}$.

\begin{lemma}\label{green}
For any Borel set $A\subset\mathbb{R}^3$,
\[-\int_A\mathcal{G}_{\bm{ u}} |\bm{x}|\di\pi_{\epsilon} =\hf[\epsilon^2] \int_{\partial A} \bm{n}\cdot\nabla|\bm{x}|\di(\pi_{\epsilon})_{\partial},\]
where $\bm{n}$ is the inward pointing unit normal to the surface $\partial A$.
\end{lemma}
\begin{proof}
Since,
\[\rho_{\epsilon} \Delta |\bm{x}| - |\bm{x}| \Delta \rho_{\epsilon} = \nabla\cdot\left(\rho_{\epsilon}\nabla|\bm{x}| - |\bm{x}|\nabla\rho_{\epsilon}\right),\]
and,
\[\nabla\cdot\left(|\bm{x}|\rho_{\epsilon} \bm{u}\right) 
 = |\bm{x}|\nabla\cdot\left(\rho_{\epsilon}\bm{u}\right) + \rho_{\epsilon} \bm{u}\cdot \nabla|\bm{x}|,\]
 it follows that,
 \[\rho_{\epsilon}\mathcal{G}_{\bm{u}} |\bm{x}| = |\bm{x}|\left(\hf[\epsilon^2]\Delta \rho_{\epsilon} -\nabla\cdot(\rho_{\epsilon}\bm{u})\right)
 +\hf[\epsilon^2]\nabla\cdot(\rho_{\epsilon} \nabla|\bm{x}|).\]
 Moreover, from equation (\ref{MF1}) it follows that
 $ 0 = \nabla\cdot(\rho_{\epsilon} \bm{v}),$
 where $\bm{v}$ is the current velocity in the stationary state. Therefore,
 \[\rho_{\epsilon}\mathcal{G}_{\bm{u}} |\bm{x}| = |\bm{x}|\left(\hf[\epsilon^2]\Delta \rho_{\epsilon} -\nabla\cdot(\rho_{\epsilon}\bm{b})\right)
 +\hf[\epsilon^2]\nabla\cdot(\rho_{\epsilon} \nabla|\bm{x}|) =\hf[\epsilon^2]\nabla\cdot(\rho_{\epsilon} \nabla|\bm{x}|) ,\]
 and the result follows by 
Green's theorem.
\end{proof}

\begin{lemma}\label{estimate}
There exists a constant $r_1>0$ such that for all $\bm{x}$ with $|\bm{x}|>r_1$,
\[ \mathcal{G}_{\bm{u}}|\bm{x}|\le-\frac{\epsilon^2\tilde{C}}{2}<0,\]
where $\tilde C = (\mu-\epsilon^2\lambda C)/(\epsilon^2\lambda)$ and $C$ is as defined in equation (\ref{assumption}).
\end{lemma}

\begin{proof}
Clearly,
\[\mathcal{G}_{\bm{u}}|\bm{x}| = \frac{\epsilon^2}{2|\bm{x}|}\left( 2+2\nabla R_{\epsilon}\cdot\bm{x}+\nabla\ln T\cdot\bm{x}\right).\]
Moreover, from Lemma \ref{alltheRS},
\[ \lim_{|\bm{x}|\rightarrow \infty} \alpha=1,\quad \lim_{|\bm{x}|\rightarrow \infty} \beta=0,\]
and so,
\[\lim\limits_{|\bm{x}|\rightarrow\infty}\frac{\epsilon^2}{|\bm{x}|}\left( 1+\nabla R_{\epsilon}\cdot\bm{x}\right) = -\frac{\mu}{\lambda}.\]
Therefore, there exists $r_1>r_0$  such that for all $\bm{x}$ with $|\bm{x}|>r_1$,
\[\frac{\epsilon^2}{|\bm{x}|}\left( 1+\nabla R_{\epsilon}\cdot\bm{x}\right)\le -\frac{\mu}{2\lambda},\]
which with the assumption in equation (\ref{assumption}) gives the result.
\end{proof}

\begin{lemma}\label{first} For every  $g\in \mathscr{L}_0(B_{r_1}^c),$
\[\int g^2 \di\pi_{\epsilon}\le \frac{4}{\tilde{C}^2}\int |\nabla g|^2\di \pi_{\epsilon}.\]
\end{lemma}

\begin{proof}
Combining Lemmas \ref{green} and \ref{estimate},
\begin{equation}\label{gtilde2}
\hf[\epsilon^2\tilde{C}]\pi_{\epsilon}(A)\le  -\int_A\mathcal{G}_{\bm{u}} |\bm{x}|\di \pi_{\epsilon}
=\hf[\epsilon^2]\int_{\partial A} \bm{n}\cdot \nabla |\bm{x}| \di (\pi_{\epsilon})_{\partial}\le  \hf[\epsilon^2](\pi_{\epsilon})_\partial(\partial A),
\end{equation}
where $\bm{n}$ is the inward unit normal vector field of $\partial A$. It follows from equation (\ref{gtilde2}) that for all $A\subset B_{r_1}^c$,
\begin{equation}\label{C}
 \frac{(\pi_{\epsilon})_{\partial}(\partial A)}{\pi_{\epsilon}(A)}\ge \tilde{C}>0.
\end{equation}
Therefore,  using Cheeger's inequality (Theorem \ref{cheeger})  on $M=B_{r_1}^c$ with a Dirichlet boundary condition on $\partial M = \left\{\bm{x}:\, |\bm{x}|=r_1\right\}$  with equation (\ref{C}) gives for sufficiently smooth $g:B_{r_1}^c\rightarrow\mathbb{R}$ such that $g(x) =0$ for $|x| = r_1$, 
\[ \frac{ \left(\int |\nabla g|^2 \di \pi_{\epsilon} \right)}{\left(\int g^2 \di \pi_{\epsilon}\right)}  \ge \frac{h^2}{4}\ge \frac{\tilde{C}^2}{4},\]
giving the result for all $g\in \mathscr{L}_0(B_{r_1}^c)$. \end{proof}

\begin{lemma}\label{lemma2} There exists a constant $r_2$ independent of $\epsilon$ with $r_2>r_1$ such that,
\[\pi_{\epsilon}(B_{r_2}^c)\le \frac{1}{4}.\]
\end{lemma}
\begin{proof}
It follows from equation (\ref{C}) that,
\[-\frac{\di}{\di r}\ln\left(\pi_{\epsilon}(B_r^c)\right) = \frac{(\pi_{\epsilon})_{\partial}(\partial B_r^c)}{\pi_{\epsilon}(B_r^c)}\ge\tilde{C},\]
and so integrating,
\[
\pi_{\epsilon}(B_{r}^c)\le \pi_{\epsilon}(B_{r_1}^c)\exp\left(-\tilde{C}(r-r_1)\right)
\le \exp\left(-\tilde{C}(r-r_1)\right),
\]
gives the result.
\end{proof}
\begin{theorem} \label{outside}
For any $ f\in C_0^1(\mathbb{R}^d)$,
\[\int_{B_{r_3}^c}f^2\di\pi_{\epsilon} \le \frac{8 }{\tilde{C}^2} \int|\nabla f|^2 \di\pi_{\epsilon} +\frac{1}{8} \int f^2 \di\pi_{\epsilon},\]
where $r_3 = r_2+ 8/\tilde{C}$.
\end{theorem}
\begin{proof}
For some constant $\delta>0$, define,
\[g(\bm{x}):= \left\{\begin{array}{ll}
0, & |\bm{x}| \le r_2,\\
\left(\frac{|\bm{x}|-r_2}{\delta}\right) f(\bm{x}),\qquad& r_2<|\bm{x}|\le r_2+\delta,\\
f(\bm{x}), & r_2+\delta<|\bm{x}|,
\end{array}\right.
\]
where $r_2$ is as in Lemma \ref{lemma2}.

It follows that as $r_2>r_1$ then $g\in\mathscr{L}_0(B_{r_1}^c)$ and so from Lemma \ref{first},
\[\int_{B_{r_2+\delta}^c}f^2\di\pi_{\epsilon}= \int_{B_{r_2+\delta}^c}g^2\di\pi_{\epsilon} \le \int g^2 \di\pi_{\epsilon}\le \frac{4}{\tilde{C}^2}\int|\nabla g|^2\di\pi_{\epsilon}.\]
Moreover, from the definition of $g$,
\[0\le|\nabla g|^2 \le 2|\nabla f|^2 + \frac{2}{\delta^2}f^2,\]
and so,
\[\int_{B_{r_2+\delta}^c}f^2\di\pi_{\epsilon} \le \frac{8}{\tilde{C}^2} \int|\nabla f|^2 \di\pi_{\epsilon} +\frac{8 }{\tilde{C}^2\delta^2} \int f^2 \di\pi_{\epsilon}.\]
Finally choosing $\delta = 8/\tilde{C}$ gives the result.
\end{proof}

We now prove a similar result on the interior of the ball using the Poincar\'{e} inequality on Euclidean balls.
\begin{theorem}\label{inside} For all $f\in C^1(\mathbb{R}^d)$ with $\int f \di\pi_{\epsilon} =0$,
\[\int_{B_{r_3} } f^2 \di\pi_{\epsilon}\le \frac{4 r_3^2}{\pi^2}\exp\left(\frac{k_0}{\epsilon^2}(r_3+1)\right) \int|\nabla f|^2\di \pi_{\epsilon}+
 \frac{1}{3}\int f^2 \di \pi_{\epsilon}.\]
\end{theorem}
\begin{proof}

Recall $V_{\epsilon} = 2 R_{\epsilon} + \ln T$ so by Lemma \ref{alltheRS}, there exists a constant $k_0>0$ such that for any $r>0$,
\begin{equation}\label{Vest}|V_{\epsilon}(\bm{x})|\le \frac{k_0}{2\epsilon^2}(|\bm{x}|+1),\quad \bm{x}\in B_r,\quad \epsilon\in (0,1].
\end{equation}
Clearly,
\[\int_{B_r} f^2\di\pi_{\epsilon} - \frac{1}{\pi_{\epsilon}(B_r)}\left(\int_{B_r}f\di\pi_{\epsilon}\right)^2  = \inf\limits_{c\in\mathbb{R}} \int_{B_r}(f-c)^2 \di\pi_{\epsilon},\]
and since $\di \pi_{\epsilon} = \exp(V_{\epsilon})\di \bm{x}$, it follows from equation (\ref{Vest}) and Theorem \ref{poincare} that,
\begin{eqnarray*}
 \inf\limits_{c\in\mathbb{R}} \int_{B_r}(f-c)^2 \di\pi_{\epsilon}
 & \le & \exp\left(\frac{k_0}{2\epsilon^2}(r+1)\right)\inf\limits_{c\in\mathbb{R}} \int_{B_r}(f-c)^2 \di\bm{x}\\
 & \le &\frac{4r^2}{\pi^2}  \exp\left(\frac{k_0}{2\epsilon^2}(r+1)\right) \int_{B_r}|\nabla f|^2 \di\bm{x}\\
  & \le & \frac{4r^2}{\pi^2} \exp\left(\frac{k_0}{\epsilon^2}(r+1)\right)\int_{B_r}|\nabla f|^2 \di\pi_{\epsilon}.
 \end{eqnarray*}
Therefore,
\begin{equation}\label{5}
\int_{B_r} f^2 \di \pi_{\epsilon} \le  \frac{4 r^2}{\pi^2} \exp\left(\frac{k_0}{\epsilon^2}(r+1)\right)
\int_{B_{r}}|\nabla f |^2\di \pi_{\epsilon}+ \frac{1}{\pi_{\epsilon}(B_{r})}\left(\int_{B_{r}} f \di\pi_{\epsilon}\right)^2.
\end{equation}
It follows from $\int f\di \pi_{\epsilon}=0$ that,
\[\left(\int_{B_{r_3}}f \di\pi_{\epsilon}\right)^2 = \left(\int_{B_{r_3}^c}f \di\pi_{\epsilon}\right)^2,\]
and so by the Cauchy-Schwartz inequality and Lemma \ref{lemma2},
\[ \left(\int_{B_{r_3}}f \di\pi_{\epsilon}\right)^2\le \left( \int_{B_{r_3}^c}\di\pi_{\epsilon}\right)\left( \int_{B_{r_3}^c}f^2\di\pi_{\epsilon}\right)\le\frac{1}{4} \int f^2 \di\pi_{\epsilon}.\]
Therefore, by (\ref{5})   we obtain,
\[\int_{B_{r_3} } f^2 \di\pi_{\epsilon}\le \frac{4r_3^2}{\pi^2}\exp\left(\frac{k_0}{\epsilon^2}(r_3+1)\right)\int|\nabla f|^2 \di\pi_{\epsilon}+
 \frac{1}{3}\int f^2\di\pi_{\epsilon},\]
which gives the result.
\end{proof}
We can now combine these results  to give a single Poincar\'{e} inequality.

\begin{cor}\label{mainresult}
For any $ f\in C_0^1(\mathbb{R}^d)$ such that $\int f\di\pi_{\epsilon}=0$ there exists a constant $k$, such that,
\[\int f^2 \di\pi_{\epsilon}\le \exp\left(k/{\epsilon}^2\right) \int |\nabla f|^2\di\pi_{\epsilon}.\]
\end{cor}
\begin{proof}
Follows directly from Theorem \ref{outside} and Theorem \ref{inside}.
\end{proof}
All that remains now is to show that this Poincar\'{e} inequality leads to the spectral gap.

\begin{theorem} There exists a constant $\gamma=\gamma(\lambda,\mu,e,C)>0$ such that the generator of the limiting diffusion process $\bm{X}_{\epsilon}$,
\[\mathcal{G}:= \hf[\epsilon^2] \Delta +\bm{b}\cdot\nabla,\]
 has a spectral  gap where,
\[\mathrm{gap}(\mathcal{G})\ge \exp\left(-\gamma/\epsilon^2\right),\] for any $\epsilon\in (0,1].$
\end{theorem}

\begin{proof} 
Let  $\lambda_1=\inf ( \sigma(-\mathcal{G})\setminus \{0\})$ where $\sigma(\cdot)$ denotes the spectrum of a linear operator. Then there exists a sequence of functions $\{f_n\}_{n\ge 1}$ such that,
\[\int f_n \di\pi_{\epsilon} = 0,\quad \int f_n^2 \di\pi_{\epsilon} = 1,\]
and,
\[\lim\limits_{n\rightarrow\infty} \int (\lambda_1 f_n + \mathcal{G} f_n)^2 \di\pi_{\epsilon} = 0.\]
By the Cauchy-Schwartz inequality,
\begin{eqnarray*}
\int(\lambda_1 f_n + \mathcal{G} f_n)^2 \di\pi_{\epsilon} & = & \left(\int  f_n^2  \di\pi_{\epsilon}\right)\left(\int(\lambda_1 f_n + \mathcal{G} f_n)^2 \di\pi_{\epsilon} \right)
\\
& \ge & 
\left(\int(\lambda_1 f_n^2 + f_n\mathcal{G} f_n) \di\pi_{\epsilon}\right)^2\ge 0.
\end{eqnarray*}
Therefore,
\[\lambda_1 =- \lim\limits_{n\rightarrow\infty}
 \int f_n\mathcal{G} f_n \di\pi_{\epsilon}.\]
Now, for $f\in\mathcal{D}(\mathcal{G})$,
\begin{eqnarray*}
-\int f\,\mathcal{G} f \di\pi_{\epsilon} & = & -\int f\left(\frac{\epsilon^2}{2}\Delta f + \bm{b}\cdot\nabla f\right)\rho_{\epsilon} \di\bm{x} \\
 & = & -\hf\int \left(\hf[\epsilon^2] \Delta f^2 -\epsilon^2 |\nabla f|^2+ \bm{b}\cdot\nabla f^2\right)\rho_{\epsilon} \di\bm{x} \\
 & = & -\hf\int \rho_{\epsilon}\, \mathcal{G}f^2 \di\bm{x} +\hf[\epsilon^2]\int |\nabla f|^2\di\pi_{\epsilon}\\
  & = & -\hf\int f^2 \, \mathcal{G}^*\rho_{\epsilon}\di\bm{x} +\hf[\epsilon^2]\int |\nabla f|^2\di\pi_{\epsilon}.
\end{eqnarray*}
Therefore, as $\rho_{\epsilon}$ is the density for the invariant measure,
\[ -\int f\,\mathcal{G} f \di\pi_{\epsilon} =  \hf[\epsilon^2]\int |\nabla f|^2\di\pi_{\epsilon}.\]
It follows that,
\[\lim\limits_{n\rightarrow\infty} \hf[\epsilon^2]\int |\nabla f_n|^2 \di\pi_{\epsilon} = \lambda_1.\]
However, from Corollary \ref{mainresult},
\[1 = \lim\limits_{n\rightarrow\infty}
\int f_n^2 \di\pi_{\epsilon} \le \exp(k/\epsilon^2)\lim\limits_{n\rightarrow\infty} \int |\nabla f_n|^2 \di\pi_{\epsilon} =\frac{ 2\lambda_1}{\epsilon^2} \exp(k/\epsilon^2).\]
Thus $\lambda_1\ge\exp(-\gamma/\epsilon^2)$ for some constant $\gamma>0$ and all $\epsilon\in(0,1]$.
 \end{proof}

 \section{Conclusion}
 We have shown that in the three dimensional case for small $\epsilon>0$, the limiting Nelson diffusion derived from the correspondence limit of the atomic elliptic state has an invariant measure which is concentrated in a neighbourhood of the Kepler ellipse in the plane $z=0$. We have also shown that the density of the limiting diffusion will converge in the infinite time limit to this invariant measure irrespective  of the eccentricity of the Kepler ellipse. We can therefore conclude that in the infinite time limit the motion of a diffusing particle will be confined to a neighbourhood of the $z=0$ plane. Hence the results of \cite{Divine} become highly relevant in establishing that this motion becomes Keplerian motion on the ellipse in the infinite time limit. We hope to investigate the ramifications of this result for planetesimal diffusions  \cite{MR782979, MR960159, MR1020069} in a future publication.

\section*{Acknowledgements}
It is a pleasure for AN and FYW to thank the Welsh Institute for Mathematical and Computational Sciences (WIMCS) for their financial support in this research. AT would also like to thank Istv\'{a}n Gy\"{o}ngy and David Elworthy for helpful conversations.

\def\cprime{$'$}

\end{document}